# Comment on the paper "Quantum mechanics needs no consciousness", by Yu and Nikolic (2011)

Catherine M Reason

Correspondence: CMRneuro@Gmail.com

This is a brief comment on the paper "Quantum mechanics needs no consciousness" by Shan Yu and Danko Nikolic [1]. Yu and Nikolic argue that the "consciousness causes collapse hypothesis" interpretation of quantum mechanics, or CCCH, can be falsified by a particular experimental setup. This claim is incorrect and the cause of the error appears to be a confusion over where and when a collapse can be assumed to occur.

The apparatus described by Yu and Nikolic is a stripped down modification of the Delayed Choice Quantum Eraser experiment designed by Kim et al [2]. In the Yu and Nikolic setup the interferometer arrangement on the idler side of the DCQE apparatus is removed, with the resulting setup as follows. Single photons from a laser pass through a double slit and are put into a superposition of paths. Each path hits a particular region on a non-linear BBO crystal and produces an entangled pair of photons by parametric down conversion. One photon from each pair, the signal photon, is directed through a lens to a detector $D_0$ which moves through the focal plane of the lens. The other photon from each pair, the idler photon, is either detected independently by a detector whose position is correlated with slit position, or allowed to disappear into the distance. (For a fuller explanation and diagram of the

experimental setup, please consult [1].)

It is clear that under such circumstances the rules of quantum mechanics predict that no interference pattern will be found at the signal detector site $D_0$, and this the authors acknowledge. Nonetheless the authors also claim the CCCH predicts that an interference pattern should be found at $D_0$, and the absence of such interference is claimed to falsify the CCCH. Yu and Nikolic are not entirely clear in their paper what sort of interference they expect; De Barros and Oas [3] point out that the DCQE apparatus is a fourth-order interference setup, which requires coincidence counting between the idler and signal detections to detect interference They also point out that the particular setup proposed by Yu and Nikolic does not actually produce fourth-order interference. Correspondence with Dr Yu has confirmed to me that the interference pattern he and Dr Nikolic are referring to is however a second-order interference pattern ( or "standard Young's double-slit interference") which would be detectable without coincidence-counting between idler and signal detections.

This is all rather puzzling and I have recently undertaken a somewhat extensive correspondence with Dr Shan Yu to try and clarify the matter. In particular, why are the authors claiming that the CCCH should predict something so at odds with the basic rules of quantum mechanics? The crux of the problem seems to be Dr Yu's belief that the very existence of "which-path" information in a quantum system, and the concomitant lack of interference, is itself a definitive indicator of wavefunction collapse (personal communication). In what follows I hope to show that this belief is erroneous for quite simple and straightforward reasons.

In order to do so it is necessary to look a little more closely at the role played by wavefunction collapse in those interpretations of quantum mechanics which incorporate it. Yu and Nikolic correctly point out that the wavefunction collapse is postulated in order to reduce a physical state which can be represented as a superposition of eigenstates in some basis to a single eigenstate. Empirically, however, it is not always so easy to distinguish between pure states and statistical mixtures -- that is to say, between physical states which are quantum superpositions on the one hand, and physical states which are mixtures of well-defined classical states on the other. In practice, one tends to distinguish between the two by the presence or absence of interference effects. If we denote the two photons in the Yu/Nikolic setup by *a* and *b*, and the two slit-positions by *L* and *R*, then the wavefunction of the two-photon system can be written in terms of:

$|aL\rangle|bL\rangle + |aR\rangle|bR\rangle$

If the system is fully entangled (as it is in the Yu and Nikolic setup) then the terms in the superposition are clearly orthogonal, and hence there will be no second-order interference at $D_0$. However the system remains technically a superposition, since the wavefunction will itself be an eigenstate of some observable which could, in principle, be measured [4]; and such a measurement would reveal an interference effect.

The first point to note here is that the second-order interference pattern disappears at $D_0$ *whether or not* one assumes a collapse has taken place, since the basic unitary evolution of the wavefunction is sufficient by itself to eliminate the interference -- wavefunction collapse is not necessary to do this. The second point to make is that, even if one endorses an interpretation of quantum mechanics which incorporates wavefunction collapse, there is still no reason to assume that wavefunction collapse has occurred *at this point in the evolution of the system*, since the particular measurement which would distinguish between a pure state and a statistical mixture has not yet been made (and indeed, such a measurement will become impractical once the wavefunction has become entangled with the environment to such an extent that it cannot be reproduced). Therefore it cannot be true to say, as Dr Yu has said to me (personal communication) that the absence of second-order interference at $D_0$ is the definitive indicator of wavefunction collapse. The collapse, if it occurs at all, could take place at any time prior to the conscious perception of some observable of the entire system by some observer; and that entire system would include the two photons, any measuring instruments, and possibly also the brains of the observers. (This is because it is only the conscious perception of a single determinate state which provides any evidence that a collapse has occurred at all.) The claims made by Yu and Nikolic, that the CCCH predicts a second-order interference pattern at $D_0$ whenever "which-path" information is available but not consciously observed, are therefore unfounded. Their claim to have falsified the CCCH is therefore incorrect.